\documentclass[11pt,bm,aps,nofootinbib,amsfonts,amssymb,preprintnumbers,superscriptaddress]{revtex4}

 \usepackage{here}

\usepackage[dvipdfmx]{graphicx}
\usepackage{amssymb,amsfonts,amsmath,bm,color,multirow,cases,empheq,hyperref}
\usepackage{mathrsfs}

\usepackage{enumerate}
\usepackage{ulem}
\usepackage{afterpage}
\newcommand{\cL}{{\cal L}}
\newcommand{\cF}{{\cal F}}

\hypersetup{%
 setpagesize=false,
 bookmarksnumbered=true,%
 bookmarksopen=true,%
 colorlinks=true,%
 linkcolor=blue,%
 citecolor=red}
\begin{document}

%%%%%%%%%%%%%%%%%%%%%%%%%%%%%%%%%%%%%%%%%%%%
%%%%%%%%%%%%%%%%%%%%%%%%%%%%%%%%%%%%%%%%%%%%
%%%%%%%%%%%%%%%%%%%%%%%%%%%%%%%%%%%%%%%%%%%%

%   << title page >>

%%%%%%%%%%%%%%%%%%%%%%%%%%%%%%%%%%%%%%%%%%%%
%%%%%%%%%%%%%%%%%%%%%%%%%%%%%%%%%%%%%%%%%%%%
%%%%%%%%%%%%%%%%%%%%%%%%%%%%%%%%%%%%%%%%%%%%

\title{Existence conditions of nonsingular dyonic black holes \\ in nonlinear electrodynamics}

\author{ Ren Tsuda}
\thanks{Corresponding author}
\email{tsuda.ren@p.chibakoudai.jp}
\affiliation{Student Support Center, Chiba Institute of Technology\\
Shibazono 2-1-1, Narashino 275-0023, Japan}

\author{ Ryotaku Suzuki}
\email{suzuki.ryotaku@nihon-u.ac.jp}
\affiliation{Laboratory of Physics, College of Science and Technology, Nihon University\\
Narashinodai 7-24-1, Funabashi 274-8501, Japan}
\affiliation{Mathematical Physics Laboratory, Toyota Technological Institute\\
Hisakata 2-12-1, Nagoya 468-8511, Japan}

\author{Shinya Tomizawa}
\email{tomizawa@toyota-ti.ac.jp}
\affiliation{Mathematical Physics Laboratory, Toyota Technological Institute\\
Hisakata 2-12-1, Nagoya 468-8511, Japan}

\date{\today}

\preprint{TTI-MATHPHYS-22}

%%%%%%%%%%%%%%%%%%%%%%%%%%%%%%%%%%%%%%%%%%%%
%%%%%%%%%%%%%%%%%%%%%%%%%%%%%%%%%%%%%%%%%%%%
%%%%%%%%%%%%%%%%%%%%%%%%%%%%%%%%%%%%%%%%%%%%

%   << abstract >>

%%%%%%%%%%%%%%%%%%%%%%%%%%%%%%%%%%%%%%%%%%%%
%%%%%%%%%%%%%%%%%%%%%%%%%%%%%%%%%%%%%%%%%%%%
%%%%%%%%%%%%%%%%%%%%%%%%%%%%%%%%%%%%%%%%%%%%

\begin{abstract} 
We study the conditions under which nonsingular black hole solutions can exist in General relativity coupled to nonlinear electrodynamics,
focusing on Lagrangians that depend on two electromagnetic invariants ${\cal F}$ and ${\cal G}$.
In particular, we derive a necessary criterion that the Lagrangian ${\cal L} ({\cal F} , {\cal G})$ must satisfy for static, spherically symmetric dyonic black hole solutions to exist without curvature singularities.
We demonstrate that this criterion is consistent with known no-go theorems for one-parameter Lagrangians ${\cal L} (\cal F)$, while allowing for new possibilities in two-parameter models.
As an explicit example, we construct a simple Lagrangian that satisfies the criterion and supports a nonsingular dyonic configuration.
We also present several additional models that satisfy the criterion, which may serve as possible candidates for supporting nonsingular dyonic solutions.
\end{abstract}
\date{\today}

\maketitle

%%%%%%%%%%%%%%%%%%%%%%%%%%%%%%%%%%%%%%%%%%%%
%%%%%%%%%%%%%%%%%%%%%%%%%%%%%%%%%%%%%%%%%%%%
%%%%%%%%%%%%%%%%%%%%%%%%%%%%%%%%%%%%%%%%%%%%

%        << Sec.1 Introduction >>

%%%%%%%%%%%%%%%%%%%%%%%%%%%%%%%%%%%%%%%%%%%%
%%%%%%%%%%%%%%%%%%%%%%%%%%%%%%%%%%%%%%%%%%%%
%%%%%%%%%%%%%%%%%%%%%%%%%%%%%%%%%%%%%%%%%%%%

\section{Introduction}

\label{sec:intro}

One of the most fundamental difficulties in general relativity is the inevitable appearance of spacetime singularities in black hole solutions. 
According to the singularity theorems proved by Penrose and Hawking in 1970 \cite{Penrose:1964wq, Hawking:1970zqf},  
under general conditions, namely the existence of matters satisfying the energy conditions, gravitational collapse leads to regions where curvature invariants diverge, and the classical description of spacetime breaks down.
However, many researchers believe that such singularities are products of the classical theory and will be removed by quantum theory of gravity in the future.
Thus, resolving such singularities has become a major goal in theoretical physics, offering insights into the nature of gravity, matter, and spacetime at the most fundamental level.

A promising approach to addressing this problem is to modify the matter content coupled to gravity, rather than to alter the gravitational framework itself. 
In particular, nonlinear electrodynamics (NED) has emerged as compelling means. 
Originally introduced by Born and Infeld in the 1930s to regularize the infinite self-energy of point charges \cite{Born:1934gh}, NED provides a nonlinear generalization of Maxwell's theory. 
When coupled to gravity, these theories yield modified energy-momentum tensors that can support spacetimes free from curvature singularities
\footnote{
Throughout this paper, we treat the nonlinear electrodynamics as a classical field that locally violates the energy conditions in the strong-field regime. As in many previous studies on nonsingular black holes, such a local violation can be regarded as effectively encoding quantum effects within a semi-classical description.
}.
Importantly, many NED models are constructed to recover Maxwell theory in the weak-field limit, 
ensuring consistency with classical physics in asymptotic regions.

The utility of this approach was demonstrated in Bardeen's 1968 solution \cite{Bardeen:1968aa}, 
which presented a nonsingular black hole metric for an asymptotically flat, static spherically symmetric spacetime without specifying the physical origin of its matter content.
Here, ``nonsingular'' means that spacetime has no curvature singularities both inside and outside the event horizon, not the absence of the horizon. 
Decades later, Ay\'{o}n-Beato and Garc\'ia provided a significant reinterpretation of the Bardeen black hole 
as an exact solution of Einstein's equations sourced by a nonlinear electromagnetic field \cite{Ayon-Beato:1998hmi, Ayon-Beato:2000mjt}.
This breakthrough initiated a systematic study of regular black hole solutions within the NED framework.
Numerous models have since been proposed.
Hayward also proposed another nonsingular black hole model with the same symmetry and asymptotics to resolve the information-loss paradox \cite{Hayward:2005gi}.
Furthermore, Fan and Wang constructed a wide class of nonsingular black hole generalizing the Bardeen black hole and the Hayward black hole \cite{Fan:2016hvf}.
These are typically characterized by one-parameter Lagrangians
$ {\cal L} \left( {\cal F} \right) $ with ${\cal F} = F_{\mu \nu} F^{\mu \nu}$, that support nonsingular black holes with purely magnetic charge. Near the center, such solutions often approach de Sitter-like geometries and exhibit finite curvature invariants.

In addition to these nonsingular black hole models, more recently proposed formulations of NED have been applied to the construction of black hole solutions. 
For example, Gullu and Mazharimousavi introduced a class of models known as double-logarithmic nonlinear electrodynamics (DLNED) \cite{Gullu:2020qni,Gullu:2020ant}, 
which admit black hole solutions with either electric or magnetic charge. 
In the magnetic case, exact solutions have been obtained, 
and their thermodynamic properties have been examined in detail. 
While these solutions are not nonsingular, such studies motivate further investigation into conditions under which NED can support black holes without singularities.

A particularly important motivation for considering magnetic charges comes from Dirac's famous 1931 proposal.
To explain the quantization of electric charge, Dirac introduced the concept of a magnetic monopole.
He demonstrated that if a single magnetic monopole exists, the product of the electric charge $e$ and magnetic charge $g$ must satisfy the quantization condition $eg = 2\pi n\hbar$ with $n \in \mathbb{Z}$.
This result provides a natural explanation for why all observed electric charges are discrete.
Although magnetic monopoles have not yet been experimentally observed, Dirac's idea has had an impact on theoretical physics.
It has motivated the study of gauge theories with topological structure and continues to inspire investigations into systems involving both electric and magnetic charges,
such as dyonic configurations in gravitational theories.

In contrast to the success of the discovery of pure magnetic nonsingular black holes, a dyonic black hole solution with both a magnetic charge $Q_{\rm m}$ and an electric charge $Q_{\rm e}$ has not been found so far,
for it was proved in Refs.~\cite{Bokulic:2022cyk, Bronnikov:2022ofk} that the one-parameter Lagrangian $ {\cal L} \left( {\cal F} \right) $ cannot have a solution with a regular center in a dyonic case.
Moreover, the electric-magnetic duality of Maxwell theory is broken in nonlinear models \cite{Gullu:2020qni}, further complicating attempts to generalize magnetically regular solutions to dyonic configurations.

To overcome these limitations, recent work has begun to consider more general two-parameter NED Lagrangians ${\cal L} \left( {\cal F} , {\cal G} \right)$, where ${\cal G} = F_{\mu\nu} * F^{\mu\nu}$ is the topological scalar invariant involving the Hodge dual of the field strength. 
Some special cases of such Lagrangians have been investigated \cite{Bokulic:2022cyk, Bronnikov:2022ofk}
which are summarized in Table \ref{tab:nonexistence},
but a general criterion for the existence of nonsingular dyonic black hole solutions in this two-parameter setting has not been established.

\begin{table}[t]
\begin{tabular}{|c|c|c|c|}
\hline
${\cal L} \left( {\cal F} , {\cal G} \right)$ & $Q_{\rm e}$ & $Q_{\rm m}$ & $Q_{\rm e} , Q_{\rm m}$ \\ \hline\hline
${\cal L} \left( {\cal F} \right)$ & Yes$^1$ & Yes & No \\
${\cal F} + \eta \left( {\cal G} \right)$, $ \forall \eta \left( {\cal G} \right)$ & ? & No & No \\
${\cal F} + a {\cal F}^s {\cal G}^u$, $a \neq 0 \in \mathbb{R}$, $s,u > 1 \in \mathbb{N} $ & ? & ? & No \\
${\cal F} + a {\cal F}^2 + b {\cal F} {\cal G} + c {\cal G}^2$, $a , b , c \in \mathbb{R}$ & ? & ? & No \\ \hline
\end{tabular} \\
\footnotesize{$^1$ The explicit form of the Lagrangian is irrelevant.}
\label{tab:nonexistence}
\caption{Summary of nonexistence theorems for electric, magnetic, and dyonic black holes.}
\end{table}

In this paper, we address this problem by deriving a necessary condition on the asymptotic behavior of the Lagrangian derivatives ${\cal L}_{\cal F}$ and ${\cal L}_{\cal G}$ at the strong field limit ${\cal F} \to \infty$ 
for static, spherically symmetric nonsingular dyonic black hole solutions to exist.
Our analysis is based on the near origin behavior of the field equations, 
and ensures the finiteness of curvature invariants such as $R$, $R_{\mu \nu} R^{\mu \nu}$, and $ R_{\mu \nu \rho \sigma} R^{\mu \nu \rho \sigma} $.
We further demonstrate that this criterion is consistent with existing no-go theorems for one-parameter ${\cal L} ({\cal F})$ models 
and provide explicit examples of two-parameter Lagrangians that satisfy it.
Notably, we show that magnetically regular solutions in ${\cal L} ({\cal F})$ theories can be extended 
to dyonic configurations in $ {\cal L} ({\cal F}, {\cal G}) $ theories under specific charge-matching conditions.

Our results also touch on broader theoretical themes. 
For instance, the inclusion of ${\cal G}$-dependent terms may affect the global structure of solutions while leaving local dynamics unchanged due to their topological nature. 
While it is essential for physical models to respect causality and unitarity constraints in general \cite{Shabad:2011hf},
we note that the region near the center of spacetime,
where the strong-field behavior of the Lagrangian becomes relevant,
is expected to be dominated by quantum gravitational effects.
Therefore, possible violations of these constraints in that region are not regarded as problematic for our present analysis.

Beyond formal consistency, regular black holes in NED are of interest in a variety of physical contexts. 
They serve as useful toy models for exploring aspects of black hole thermodynamics, information loss, and quantum gravity phenomenology. 
Some have even been proposed as candidates for astrophysical compact objects or dark matter remnants, potentially testable via gravitational wave observations.

The remaining parts of this article are organized as follows;
in the next section, we briefly review the general relativity coupled to NED
and present the field equations for a static and spherically symmetric spacetime.
In Sec. \ref{sec:existence}, we derive the necessary condition on the Lagrangian for the existence of nonsingular black hole solutions.
Some explicit simple examples of two-parameter Lagrangian providing a nonsingular dyonic black hole are given in Sec. \ref{sec:example}.
Sec. \ref{sec:conclusion} is devoted to conclusion.

%%%%%%%%%%%%%%%%%%%%%%%%%%%%%%%%%%%%%%%%%%%%
%%%%%%%%%%%%%%%%%%%%%%%%%%%%%%%%%%%%%%%%%%%%
%%%%%%%%%%%%%%%%%%%%%%%%%%%%%%%%%%%%%%%%%%%%

%        << Sec.2 Preliminary >>

%%%%%%%%%%%%%%%%%%%%%%%%%%%%%%%%%%%%%%%%%%%%
%%%%%%%%%%%%%%%%%%%%%%%%%%%%%%%%%%%%%%%%%%%%
%%%%%%%%%%%%%%%%%%%%%%%%%%%%%%%%%%%%%%%%%%%%

\section{Preliminary}

\label{sec:preliminary}

Here we briefly review the general relativity coupled to NED
and introduce the field equations for a spherically symmetric spacetime.
The most general form of the action in the theory is given by
\begin{align}
\label{eq:action}
S = \frac{1}{16 \pi} \int d^4 x \sqrt{-g} \left[ R - 2 \Lambda - {\cal L} \left( {\cal F} , {\cal G} \right) \right],
\quad {\cal F} \equiv F_{\mu \nu} F^{\mu \nu},
\quad {\cal G} \equiv F_{\mu \nu} * F^{\mu \nu},
\end{align}
where $F_{\mu \nu} = \partial_\mu A_\nu - \partial_\nu A_\mu$ is the Faraday tensor and $*$ denotes the Hodge star operation\footnote{The Hodge dual of the Faraday tensor is given by 
$\displaystyle * F_{\mu \nu} = \frac{1}{2} \sqrt{-g} \epsilon_{\mu \nu \rho \sigma} F^{\rho \sigma}$,
where $\epsilon_{\mu \nu \rho \sigma}$ is the totally antisymmetric tensor with $\epsilon_{tr\theta\phi} = +1$.
}.
Especially when ${\cal L} \left( {\cal F} , {\cal G} \right) = {\cal F}$, it reproduces the standard action of Einstein-Maxwell theory.
From the action (\ref{eq:action}), one can derive the Einstein equations and the source free NED equations
\begin{align}
\label{eq:Eineq1}
R_{\mu \nu} - \frac{1}{2} R g_{\mu \nu} + \Lambda g_{\mu \nu} =& T_{\mu \nu}, \\
\label{eq:elemageq1}
\nabla_\mu \left( {\cal L_F} F^{\mu \nu} + {\cal L_G} * F^{\mu \nu} \right) =& 0,
\end{align}
with the energy-momentum tensor
\begin{align}
\label{eq:Tmunu}
T_{\mu \nu} = \frac{ \partial {\cal L} }{ \partial g^{\mu \nu} } - \frac{1}{2} {\cal L} g_{\mu \nu}
= 2 \left[ {\cal L_F} F_{\mu \alpha} F_{\nu}{}^\alpha + \frac{1}{4}\left({\cal G}{\cal L_G} -{\cal L}\right) g_{\mu\nu} \right],
\end{align}
where ${\cal L}_{\cal F} = \partial {\cal L} / \partial {\cal F}$ and ${\cal L}_{\cal G} = \partial {\cal L} / \partial {\cal G}$.
In addition, the Faraday tensor $F_{\mu \nu}$ satisfies the Bianchi identity
\begin{align}\label{eq:Bianchi}
\quad \nabla_{ [ \rho} F_{\mu \nu ]} = 0,\quad  {\rm or \quad } \nabla_\mu * F^{\mu\nu}=0.
\end{align}

The most general ansatz for a static and spherically symmetric spacetime is given by
\begin{align}
ds^2 = - f (r) dt^2 + \frac{dr^2}{f (r)} + r^2 \left( d \theta^2 + \sin^2 \theta d \phi^2 \right), 
\quad
A_\mu dx^\mu = a (r) dt - Q_{\rm m} \cos \theta d \phi, \label{eq:metric+A}
\end{align}
in which, ${\cal F}$ and ${\cal G}$ are written as
\begin{align}
\label{eq:calF}
{\cal F} =& - 2 \left( a' \left( r \right) \right)^2 + 2 \left( \frac{ Q_{\rm m} }{r^2} \right)^2, \\
\label{eq:calG}
{\cal G} =& 4 a' \left( r \right) \frac{ Q_{\rm m} }{r^2}.
\end{align}
Equations (\ref{eq:Eineq1}) and (\ref{eq:elemageq1}) reduce to
\begin{align}
\label{eq:Eineq2}
&\left( r^2 f' (r) \right)' = 4 r^2 \left[ {\cal L_F} \left( \frac{ Q_{\rm m} }{r^2} \right)^2 + {\cal L_G} a'(r) \frac{ Q_{\rm m} }{r^2} - \frac{1}{4} {\cal L} \right] - 2 \Lambda r^2 , \\
\label{eq:Eineq3}
&\left( r f (r) \right)' = 1 - 2 r^2 \left[ {\cal L_F} \left( a' (r) \right)^2 - {\cal L_G} a'(r) \frac{ Q_{\rm m} }{r^2} + \frac{1}{4} {\cal L}\right] - \Lambda r^2, \\
\label{eq:elemageq2}
&(r^2 {\cal L}_{\cal F} a'(r) )'=( {\cal L}_{\cal G} Q_{\rm m} )',
\end{align}
where the electric and magnetic charges, $Q_{\rm e}$ and $Q_{\rm m}$ are defined by
\begin{align}
Q_{\rm e} \equiv& \frac{1}{4 \pi} \int_S * \left( {\cal L}_{\cal F} F + {\cal L}_{\cal G} * F \right)
= \frac{1}{4 \pi} \int_S d \theta d \phi r^2 \sin \theta \left( {\cal L}_{\cal F} F^{r t} + {\cal L}_{\cal G} * F^{r t} \right) , \label{eq:Q_e}\\
Q_{\rm m} \equiv& \frac{1}{4 \pi} \int_S F
= \frac{1}{4 \pi} \int_S d \theta d \phi r^2 \sin \theta * F^{r t}. \label{eq:Q_m}
\end{align}
Note that $F = \frac{1}{2} F_{\mu \nu} dx^\mu \wedge dx^\nu$ in the second equalities of eqs. (\ref{eq:Q_e}) and (\ref{eq:Q_m}) is the symbolic expression of the Faraday tensor.
Then, eq.~(\ref{eq:elemageq2}) is integrated as
\begin{align}
r^2 {\cal L}_{\cal F} a'(r) =& {\cal L}_{\cal G} Q_{\rm m} - Q_{\rm e}.\label{eq:elemageq2-int}
\end{align}

Moreover, combining eqs. (\ref{eq:Eineq2}) and (\ref{eq:Eineq3}), we can make the following covariant quantity 
\begin{align}
\label{eq:Eineq4}
R_{\mu \nu} R^{\mu \nu} - \frac{1}{4} R^2 
= \frac{ 1 }{ 4 r^4 } \left[ r^2 f'' (r) - 2 f (r) + 2 \right]^2
= 4 {\cal L}_{\cal F}^2 \left[ \left( a'(r) \right)^2 + \left( \frac{ Q_{\rm m} }{r^2} \right)^2 \right]^2 .
\end{align}
Note that similar covariant quantities are used in Ref. \cite{Bokulic:2022cyk}.
In the coming section, we will use this to verify the regularity of spacetime, including the regularity of each curvature invariants $R, R^{\mu\nu}R_{\mu\nu}, R^{\mu\nu\rho\sigma}R_{\mu\nu\rho\sigma}$.

%%%%%%%%%%%%%%%%%%%%%%%%%%%%%%%%%%%%%%%%%%%%
%%%%%%%%%%%%%%%%%%%%%%%%%%%%%%%%%%%%%%%%%%%%
%%%%%%%%%%%%%%%%%%%%%%%%%%%%%%%%%%%%%%%%%%%%

%        << Sec.3 Existence conditions of nonsingular dyonic black holes >>

%%%%%%%%%%%%%%%%%%%%%%%%%%%%%%%%%%%%%%%%%%%%
%%%%%%%%%%%%%%%%%%%%%%%%%%%%%%%%%%%%%%%%%%%%
%%%%%%%%%%%%%%%%%%%%%%%%%%%%%%%%%%%%%%%%%%%%

\section{Existence conditions of nonsingular dyonic black holes}

\label{sec:existence}

In this section, we study the existence of nonsingular dyonic black hole solutions to eqs. (\ref{eq:Eineq2})--(\ref{eq:elemageq2}).
It was proved that, in the one-parameter Lagrangian $ {\cal L} \left( {\cal F} \right) $, nonsingular dyonic  black hole solutions cannot exist \cite{Bokulic:2022cyk, Bronnikov:2022ofk}.
However, the case of two-parameter Lagrangian ${\cal L} \left( {\cal F} , {\cal G} \right)$ is less well understood.
This problem has been partially studied in \cite{Bokulic:2022cyk, Bronnikov:2022ofk}, where constraints were obtained for some specific forms of ${\cal L} \left( {\cal F} , {\cal G} \right)$
(See also Table \ref{tab:nonexistence}).
Here, we consider dyonic solutions, $ Q_{\rm e} Q_{\rm m} \neq 0 $ in more general $ {\cal L} \left( {\cal F} , {\cal G} \right) $.

From eq. (\ref{eq:Eineq4}), the regularity at the center $r=0$ of a spacetime requires 
$\left| {\cal L}_{\cal F} \right| \left( a' \left( r \right) \right)^2 < \infty $
and
$\left| {\cal L}_{\cal F} \right| r^{-4} < \infty $, which can be written as\footnote{To avoid confusion, we specify the definitions of Landau notation.
\begin{enumerate}
\item Big-O notation: $ f(x) = {\cal O} \left( g (x) \right) $ as $x \to a$. \\
It means that $f(x)$ is bounded by a constant multiple of $g(x)$ near $x=a$ and is defined mathematically as follows:\\
There exist constants $C > 0$ and $\delta > 0$ 
   such that $ \displaystyle \left| f(x) \right| \leq C \left| g(x) \right| ~ \mbox{for all}~x~\mbox{in}~( a - \delta , a + \delta ) \backslash \left\{ a \right\} $.
\item Little-O notation: $f(x) = o \left( g(x) \right)$ as $x \to a$. \\
This notation is used to describe that $f(x)$ becomes negligible compared to $g(x)$ as $x \to a$, namely \\
$ \displaystyle \lim_{x \to a} \frac{ f(x) }{ g(x) } = 0$.
\end{enumerate}
}
\begin{align}
\label{eq:L_F}
& {\cal L}_{\cal F} = {\cal O} \left( r^{4 + \delta_1} \right) \quad \left( r \to 0 \right), \\
\label{eq:a}
& a' \left( r \right) = {\cal O} \left( r^{- \left( 2 + \delta_2 / 2 \right)} \right) \quad \left( r \to 0 \right),
\end{align}
where $ \delta_1 $ and $ \delta_2 $ are some constants satisfying $ \delta_1 \geq 0 $ and $ \delta_1 - \delta_2 \geq 0 $.
Hereafter in this section, we restrict our discussion to the case $r \to 0$.
Eq. (\ref{eq:elemageq2-int}) gives rise to
\begin{align}
{\cal L}_{\cal G} Q_{\rm m} - Q_{\rm e} = r^2 {\cal L}_{\cal F} a' \left( r \right) = {\cal O} \left( r^{4 + \delta_1 - \delta_2 / 2 } \right).
\end{align}
Then, ${\cal L} \left( {\cal F} , {\cal G} \right)$ should satisfy the following conditions
\begin{align}
\label{eq:LForder}
& {\cal L}_{\cal F} = {\cal O} \left( r^{4 + \delta_1} \right), \\
\label{eq:LGorder}
& {\cal L}_{\cal G} - \alpha =  {\cal O} \left( r^{4 + \delta_1 - \delta_2 / 2} \right),
\end{align}
where
\begin{align}
\alpha=Q_{\rm e} / Q_{\rm m}.
\end{align}
Substituting eq. (\ref{eq:a}) into eqs. (\ref{eq:calF}) and (\ref{eq:calG}), 
and assuming $ \delta_2 < -2$ so that the gauge field $a(r)$ is finite in the limit $r \to 0$,
we obtain the behaviors of ${\cal F}$ and ${\cal G}$ as
\begin{align}
& {\cal F} = {\cal O} \left( r^{-4} \right), \label{eq:Fr}\\
& {\cal G} = {\cal O} \left( r^{- ( 4 + \delta_2 / 2 )} \right),\label{eq:Gr}
\end{align}
where it should be emphasized that the divergence in eq.~(\ref{eq:Fr}) arises inevitably from the presence of the magnetic monopole at $r=0$, that is, as a direct consequence of Gauss's law for the magnetic field, whereas the divergence in eq.~(\ref{eq:Gr}) can be weakened by $-\delta_2/2\ (>1)$, reflecting a deviation from Gauss's law for the electric field in NED. 
Thus, comparing with eqs. (\ref{eq:LForder}) and (\ref{eq:LGorder}), we can read off the following criterion at the strong field limit ${\cal F} \to \infty$
\begin{align}
\label{eq:LForder2}
& {\cal L}_{\cal F} = {\cal O} \left( {\cal F}^{- (1 + \delta_1 / 4)} \right), \\
\label{eq:LGorder2}
& {\cal L}_{\cal G} - Q_{\rm e} / Q_{\rm m} = {\cal O} \left( {\cal F}^{- (1 + \delta_1 / 4 - \delta_2 / 8)} \right).
\end{align}
Indeed, this criterion consistently excludes nonsingular dyonic  black holes in the known cases (Table \ref{tab:nonexistence}).

In closing this section we comment on the curvature polynomials $R$, $R_{\mu \nu} R^{\mu \nu}$, and $R_{\mu \nu \rho \sigma} R^{\mu \nu \rho \sigma}$ to confirm the regularity of spacetime under the condition (\ref{eq:L_F}) and (\ref{eq:a}).
From eqs. (\ref{eq:Eineq2}) and (\ref{eq:Eineq3}) we can read off the relation between the metric function $f(r)$, the gauge field $a(r)$, and the two-parameter Lagrangian ${\cal L} \left( {\cal F} , {\cal G} \right)$ as
\begin{align}
\label{eq:Eineq5}
r^2 f''(r) + 2 \left( 1 - f(r) \right) = 4 r^2 {\cal L}_{\cal F} \left[ \left( a'(r) \right)^2 + \left( \frac{ Q_{\rm m} }{r^2} \right)^2 \right].
\end{align}
Eqs. (\ref{eq:L_F}) and (\ref{eq:a}), together with (\ref{eq:Eineq5}), lead to
\begin{align}
\label{eq:forder}
1 - f(r) = {\cal O} \left(r^{2 + \delta_1}\right).
\end{align}
From this, the curvature scalars at $r \to 0$ are presented as follows:
\begin{align}
R &= - f''(r) - \frac{4 f'(r)}{r} + \frac{2 \left( 1 - f(r) \right)}{r^2}
= {\cal O} \left( r^{\delta_1} \right), \\
\nonumber
R_{\mu \nu} R^{\mu \nu} &= \frac{ \left( f''(r) \right)^2}{2} 
+ \frac{2 f'(r) f''(r)}{r} 
+ \frac{4 \left( f'(r) \right)^2}{r^2} 
- \frac{ 4 \left( 1 - f(r) \right) f'(r) }{r^3}
+ \frac{ 2 \left( 1 - f(r) \right)^2 }{r^4} \\
&= {\cal O} \left( r^{ 2 \delta_1} \right), \\
R_{\mu \nu \rho \sigma} R^{\mu \nu \rho \sigma} &= \left( f''(r) \right)^2
+ \frac{ 4 \left( f'(r) \right)^2 }{r^2} 
+ \frac{ 4 \left( 1 - f(r) \right)^2 }{r^4}
= {\cal O} \left( r^{ 2 \delta_1} \right),
\end{align}
and it is obvious that these are regular at the center.

Although the action (\ref{eq:action}) includes the cosmological constant $\Lambda$, 
it does not affect the analysis above. 
The existence condition for nonsingular dyonic black holes
depends only on the near-origin behavior of the fields, 
where the influence of $\Lambda$ becomes negligible. 
Thus, our arguments hold equally well for both $\Lambda = 0$ and $\Lambda \neq 0$.

%%%%%%%%%%%%%%%%%%%%%%%%%%%%%%%%%%%%%%%%%%%%
%%%%%%%%%%%%%%%%%%%%%%%%%%%%%%%%%%%%%%%%%%%%
%%%%%%%%%%%%%%%%%%%%%%%%%%%%%%%%%%%%%%%%%%%%

%        << Sec.4 Simple examples of two-parameter Lagrangian providing a nonsingular dyonic black hole >>

%%%%%%%%%%%%%%%%%%%%%%%%%%%%%%%%%%%%%%%%%%%%
%%%%%%%%%%%%%%%%%%%%%%%%%%%%%%%%%%%%%%%%%%%%
%%%%%%%%%%%%%%%%%%%%%%%%%%%%%%%%%%%%%%%%%%%%

\section{Simple examples of two-parameter Lagrangian providing a nonsingular dyonic black hole}

\label{sec:example}

In this section, we present several simple examples of two-parameter Lagrangians $ {\cal L} \left( {\cal F} , {\cal G} \right) $ that satisfy the existence criterion derived in the previous section.
These models demonstrate how nonsingular magnetic black hole solutions known from one-parameter theories can be extended to dyonic configurations under appropriate charge-matching condition.

Note that including odd power of the scalar invariant ${\cal G}$ in the Lagrangian generally results in violation of CP symmetry, which remains unobserved in current experimental data.
In addition, one of the Lagrangian considered in this section exhibit divergence at ${\cal G} = 0$.
Nevertheless, since our concern lies on the strong-field region near the center of spacetime, such weak-field behavior is not relevant for the present analysis.

As the simplest example that satisfies the criterion (\ref{eq:LForder2}) and (\ref{eq:LGorder2}),
let us consider the Lagrangian
\begin{align}
\label{eq:aGhF}
{\cal L} \left( {\cal F} , {\cal G} \right) = \alpha {\cal G} + h \left( {\cal F} \right)
\quad \left( \alpha \neq 0 \right), 
\end{align}
where $h \left( {\cal F} \right)$ are any one-parameter Lagrangians that allows nonsingular magnetically charged black holes \cite{Ayon-Beato:1998hmi,Fan:2016hvf,Ayon-Beato:2000mjt}. 
See also \cite{Bronnikov:2022ofk} for a comprehensive review.
Then, ${\cal L_F}$ and ${\cal L_G}$ become
\begin{align}
{\cal L}_{\cal F} = \frac{\partial h}{ \partial {\cal F}} \equiv h_{\cal F} ,
\quad
{\cal L}_{\cal G} = \alpha.
\end{align}
Since the term $\alpha {\cal G}$ in the Lagrangian (\ref{eq:aGhF}) is linear in the topological scalar ${\cal G} = F_{\mu \nu} * F^{\mu \nu}$,
it does not contribute to the local dynamics.
This is because $ \sqrt{-g} {\cal G}$ can be expressed as a total derivative in four-dimensional spacetime, namely $\displaystyle \int d^4 x \sqrt{-g} {\cal G} = \int F \wedge F = \int d (A \wedge F) $,
and thus its variation with respect to the gauge potential vanishes.
Consequently, the term does not affect the equations of motion derived from the action.
(Nevertheless, such topological term may influence the global structure of the field theory.)
In fact,
we find from eq.~(\ref{eq:elemageq1}) that its existence  changes neither the Einstein equation~(\ref{eq:Eineq1}) (more precisely, the energy momentum tensor~(\ref{eq:Tmunu}) ) nor  the field equation of NED given by $h(\cF)$, 
\begin{align}
\nabla_\mu \left(h_\cF F^{\mu \nu}  \right) =& 0,
\end{align}
because  the second term in the NED equation for the Lagrangian~(\ref{eq:aGhF}),
\begin{align}
\nabla_\mu \left( h_\cF F^{\mu \nu} \right) +\alpha \nabla_\mu * F^{\mu \nu}  =& 0,
\end{align}
vanishes from the Bianchi identity~(\ref{eq:Bianchi}).
Nevertheless, under the assumptions of staticity, spherically symmetry,  such a topological term contributes to the integration of the NED equation~(\ref{eq:elemageq2}) as
\begin{eqnarray}
r^2 h_{\cal F} a'(r) =& \alpha Q_{\rm m} - Q_{\rm e}, \label{eq:a'}
\end{eqnarray}
where the first term comes from the topological term and the second term is the integration constant determined from the definition~(\ref{eq:Q_e}) of an electric charge. 
We should note that $\cF$ is given by eq.~(\ref{eq:calF}), hence, since it includes $a'(r)$ in $h_\cF(\cF)$ in the left-hand side of eq.~(\ref{eq:a'}), 
it is not easy to solve eq.(\ref{eq:a'}) and  obtain the time component $a(r)$ of the gauge potential for a dyonic black hole solution in the Lagrangian~ $\cL(\cF,\cal G)$~(\ref{eq:aGhF}).
However, it is interesting from the regularity conditions (\ref{eq:LGorder2}) of the metric that $\alpha$ must be equal to the charge ratio 
\begin{align}
\alpha = \frac{Q_{\rm e}} { Q_{\rm m}}. 
\end{align}
In particular, for purely magnetic configurations where $Q_{\rm e} = 0$, this implies $\alpha = 0$.
Thus, the Lagrangian reduces to ${\cal L} ({\cal F}) = h ({\cal F})$ as expected.
Although this is obvious, we emphasize it to clarify the correspondence between pure magnetic and dyonic solutions.

Hence, from this and eq.~(\ref{eq:a'}), we again obtain $a(r)=0$.
 One can easily see that as a result of imposing the regularity of the metric, the Einstein equations (\ref{eq:Eineq2}) and (\ref{eq:Eineq3}) reduce to those coupled with
a magnetic monopole in a one-parameter Lagrangian ${\cal L} \left( {\cal F} \right) = h \left( {\cal F} \right)$.
Therefore, a nonsingular black hole solution with a magnetic charge $Q_{\rm m}$ in ${\cal L} \left( {\cal F} \right) = h \left( {\cal F} \right)$
can also be a nonsingular black hole solution with both a magnetic charge $Q_{\rm m}$ and an electric charge $ Q_{\rm e} = \alpha Q_{\rm m} $ in ${\cal L} \left( {\cal F} , {\cal G} \right) = \alpha {\cal G} + h \left( {\cal F} \right)$.

As one of examples, let us consider the asymptotically flat, static and spherically symmetric black hole solution with a magnetic monopole given by~\cite{Fan:2016hvf}
\begin{eqnarray}
ds^2&=&-f(r)dt^2+\frac{dr^2}{f(r)}+r^2(d\theta^2+\sin^2\theta d\phi^2), \quad f(r)= 1-\frac{2M-2q^3\tilde \alpha^{-1}}{r}-\frac{2 \tilde\alpha^{-1}q^3 r^{\mu-1}}{(r^\nu+q^\nu)^{\frac{\mu}{\nu}}}, \label{eq:metricFW}
\end{eqnarray}
and
\begin{align}
 A_\mu dx^\mu= \frac{q^2}{\sqrt{2\tilde\alpha}} \cos\theta d\phi,\quad \cF = \frac{q^4}{\tilde\alpha r^4},\label{eq:F-magnetic}
\end{align}
where $M$ is the ADM mass of the spacetime. 
This can be derived from the Einstein equation coupled with the NED given by the Lagrangian  
\begin{align}
  h({\cal F})=\frac{4\mu}{\tilde\alpha}\frac{(\tilde\alpha{\cal F})^{\frac{\nu+3}{4}}}{  \left(1+(\tilde\alpha{\cal F})^{\frac{\nu}{4}} \right)^{\frac{\mu+\nu}{\nu}}  },\label{eq:LF-magnetic}
\end{align}
where $\mu,\nu$ and $\tilde\alpha$ are free parameters of the theory.
This theory reproduces the usual Maxwell theory at the weak field limit ${\cal F}\to 0$ only when $\nu=1$.
Naively, the above solution may appear to be purely magnetic because $a(r)=0$.
However, in the two-parameter nonlinear electrodynamics described by ${\cal L}({\cal F},{\cal G})$, 
the electric charge defined by Eq.~(\ref{eq:Q_e}) yields nonzero electric charge as $Q_{\rm e} = \alpha Q_{\rm m} \neq 0$ from eq.~(\ref{eq:aGhF}).
Therefore, 
from the above discussion the metric~(\ref{eq:metricFW}) and gauge potential~(\ref{eq:F-magnetic}) give a nonsingular magnetic charged black hole $(Q_ {\rm e}=0,Q_ {\rm m}\not=0)$  in Einstein theory coupled with the NED of the Lagrangian   $h(\cF)$,  whereas they give a nonsingular dyonic black hole $(Q_ {\rm e}\not=0,Q_ {\rm m}\not=0)$ in ones coupled with the NED of the Lagrangian   $\cL(\cF,{\cal G})=\alpha {\cal G}+h(\cF)$.

As the second example, let us consider the Lagrangian  
\begin{eqnarray}
\cL({\cal F},{\cal G})=\alpha {\cal G}+\ln (1+\beta {\cal G})+h({\cal F}),
\end{eqnarray}
where $h \left( {\cal F} \right)$ is the same one given in~eq.(\ref{eq:LF-magnetic}). 
From the behavior of the derivatives at $\cF,{\cal G}\to\infty$,
\begin{eqnarray}
\cL_{\cal F}=h_{\cal F}={\cal O}(\cF^{-1}),\quad \cL_{\cal G}=\alpha +\frac{\beta}{1+\beta {\cal G}}=\alpha+{\cal O}({\cal G}^{-1}),
\end{eqnarray}
the regularity condition~(\ref{eq:LGorder2}) requires $\alpha=Q_{\rm e} /Q_{\rm m}$.
From the same discussion, the metric~(\ref{eq:metricFW}) and gauge potential~(\ref{eq:F-magnetic}) for a nonsingular magnetic charged black hole $(Q_ {\rm e}=0,Q_ {\rm m}\not=0)$  in the NED of $h(\cF)$  gives a nonsingular dyonic black hole $(Q_ {\rm e}\not=0,Q_ {\rm m}\not=0)$ in NED of $\cL({\cal F},{\cal G})=h(\cF)+\alpha {\cal G}+\ln(1+\beta {\cal G})$ if one adds the solution $a(r)$ to the differential equation
\begin{eqnarray}
 r^2 h_\cF a'(r)=\frac{\beta Q_ {\rm m}}{1+\beta {\cal G}} \label{eq:log}
\end{eqnarray}
 to the gauge potential~(\ref{eq:F-magnetic}), where we should note that $\cF$ is not given by $\cF$ in eq.~(\ref{eq:F-magnetic}) but eq.~(\ref{eq:calF}), hence, it is so difficult to solve eq. (\ref{eq:log}) analytically.

\medskip
More generally, let us assume the Lagrangian has the approximate form in the strong field regime
\begin{eqnarray}
\cL({\cal F},{\cal G}) \simeq \alpha {\cal G}+\sum_{n=1}^N(\beta_n {\cal G})^{-n}+h({\cal F}), \label{eq:Lag3}
\end{eqnarray}
where $h \left( {\cal F} \right)$ is the same in~eq.(\ref{eq:LF-magnetic}). 
Then, the condition~(\ref{eq:LGorder2}) can be written as
\begin{eqnarray}
 \cL_{\cal G}-\frac{Q_{\rm e}}{Q_{\rm m}}
 =\alpha -\frac{Q_{\rm e}}{Q_{\rm m}} - \sum_{n=1}^N n \beta_n (\beta_n {\cal G})^{-n-1}=\alpha-\frac{Q_{\rm e}}{Q_{\rm m}}+{\cal O}({\cal G}^{-1}),
\end{eqnarray}
which leads to $\alpha=Q_{\rm e} / Q_{\rm m}$.
Therefore, the metric~(\ref{eq:metricFW}) and gauge potential~(\ref{eq:F-magnetic}) for a nonsingular magnetic charged black hole $(Q_ {\rm e}=0,Q_ {\rm m}\not=0)$  in NED with $h(\cF)$ also gives a nonsingular dyonic black hole $(Q_ {\rm e}\not=0,Q_ {\rm m}\not=0)$ in NED with the Lagrangian~(\ref{eq:Lag3})  if one adds the solution $a(r)$ to the differential  equation
\begin{eqnarray}
 r^2 h_\cF a'(r)= - Q_ {\rm m}\sum_{n=1}^N n \beta_n (\beta_n {\cal G})^{-n-1} \label{eq:power}
\end{eqnarray}
to the gauge potential~(\ref{eq:F-magnetic}), where, as mentioned previously, $\cF$ is not given by $\cF$ in eq.~(\ref{eq:F-magnetic}) but eq.~(\ref{eq:calF}).

\medskip

Finally, we take the following exponential type of the Lagrangian 
\begin{eqnarray}
\cL({\cal F})= {\cal F}e^{-\beta\cal F},
\end{eqnarray}
which has the Maxwell limit at the weak field $\cF\to 0$. 
It is obvious that the regularity condition~(\ref{eq:LForder2})  can be satisfied $\cF \to \infty$ due to the suppression of $e^{-\beta \cF}$ at $\cF\to\infty$. 
Therefore, this NED does not prohibit the existence of nonsingular dyonic black holes and nonsingular purely electrically or magnetically charged black holes. 
Moreover, we can  consider the two parameter Lagrangian
\begin{eqnarray}
\cL({\cal F},{\cal G})=\alpha_1{\cal G}+\alpha_2 {\cal G}e^{-\beta\cal G}+h(\cF),
\end{eqnarray}
where $h \left( {\cal F} \right)$ is given by eq.~(\ref{eq:LF-magnetic}). 
The regularity condition~(\ref{eq:LGorder2})  can written as
\begin{eqnarray}
\cL_{\cal G}-\frac{Q_{\rm e}}{Q_{\rm m}}=\alpha_1+\alpha_2 e^{-\beta {\cal G}}(1-\beta {\cal G})-\frac{Q_{\rm e}}{Q_{\rm m}}=\alpha_1-\frac{Q_{\rm e}}{Q_{\rm m}}+{\cal O}({\cal G}^{-1}),
\end{eqnarray}
which requires
\begin{eqnarray}
\alpha_1=\frac{Q_{\rm e}}{Q_{\rm m}}.
\end{eqnarray}
Thus,  in the same way as above, nonsingular dyonic black hole solutions can be obtained  if one can obtain the solution to the differential  equation
\begin{eqnarray}
 r^2 h_\cF a'(r)=Q_ {\rm m}\alpha_2 e^{-\beta {\cal G}}(1-\beta {\cal G}).
\end{eqnarray}

%%%%%%%%%%%%%%%%%%%%%%%%%%%%%%%%%%%%%%%%%%%%
%%%%%%%%%%%%%%%%%%%%%%%%%%%%%%%%%%%%%%%%%%%%
%%%%%%%%%%%%%%%%%%%%%%%%%%%%%%%%%%%%%%%%%%%%

%        << Sec.4 Conclusion >>

%%%%%%%%%%%%%%%%%%%%%%%%%%%%%%%%%%%%%%%%%%%%
%%%%%%%%%%%%%%%%%%%%%%%%%%%%%%%%%%%%%%%%%%%%
%%%%%%%%%%%%%%%%%%%%%%%%%%%%%%%%%%%%%%%%%%%%

\section{Conclusion}

\label{sec:conclusion}

In this article, we have investigated the existence of static spherically symmetric nonsingular solutions with a dyonic charge
in Einstein theory coupled to NED with a two-parameter Lagrangian $ {\cal L} \left( {\cal F} , {\cal G} \right) $.
As a result, we have obtained a simple criterion for the existence of dyonic solutions
in the strong field limit, given by eqs. (\ref{eq:LForder2}) and (\ref{eq:LGorder2}).
This criterion is consistent with previous studies \cite{Bokulic:2022cyk, Bronnikov:2022ofk}.
Moreover, we have shown that
nonsingular solutions with a magnetic charge in a Lagrangian ${\cal L} \left( {\cal F} \right) = h \left( {\cal F} \right)$ can also be interpreted as
nonsingular solutions with a dyonic charge in a two-parameter Lagrangian ${\cal L} \left( {\cal F} , {\cal G} \right) = \alpha {\cal G} + h \left( {\cal F} \right) $, which serves as the simplest example satisfying the criterion. 
Furthermore, we have presented three types of two-parameter Lagrangians originating from the one-parameter Fan--Wang model that satisfy the criterion. 
If the differential equations for the temporal component of the gauge potential can be solved, 
nonsingular dyonic black hole solutions can be obtained using the same metric and magnetic gauge potential as the Fan--Wang black holes within Einstein-NED framework.

\medskip
It should be emphasized that the criteria~(\ref{eq:LForder2}) and (\ref{eq:LGorder2}) derived in this paper provide only necessary conditions for the existence of black hole solutions with a regular center; they do not guarantee the actual existence of nonsingular solutions.
Moreover, as in many previous studies of regular black holes in nonlinear electrodynamics, our analysis has been restricted to the elimination of curvature singularities at the center $r = 0$.
Therefore, these criteria do not exclude the possibility of curvature singularities appearing outside the event horizon or away from the center inside the horizon.
In addition, the criteria do not even ensure the existence of an event horizon itself.
Nevertheless, when one attempts to construct exact solutions that possess an event horizon and are free from singularities everywhere both inside and outside it, the conditions~(\ref{eq:LForder2}) and (\ref{eq:LGorder2}) can serve as a useful diagnostic tool.
This is because, within any nonlinear electrodynamics theory that fails to satisfy these conditions, such nonsingular black hole solutions can never exist, and thus the corresponding nonlinear electrodynamics can be excluded from consideration.
As a future direction, we also aim to extend the present criteria to incorporate the existence of an event horizon.

\medskip
It is also interesting to consider the extension of the existence condition to rotating cases.
Up to now, although various nonsingular rotating  black hole models have been proposed \cite{Torres:2022twv,Takeuchi:2016nrj,Lima:2023jtl}, 
none of them have been obtained as exact solutions of any known field theory, including NED.
It will be useful to study what kind of metric ansatz and matter fields allow or forbid the existence of nonsingular rotating black hole solutions.
Even though not conclusive, existence conditions of rotating solutions were discussed for one-parameter Lagrangians in Ref. \cite{Dymnikova:2015hka}.

\medskip
The thermodynamic properties of nonsingular black holes also deserve investigation.
It is important to determine whether the known laws of black hole thermodynamics can apply to such nonsingular configurations.
Moreover,  it is worth exploring whether singularities can be removed even in non-stationary solutions of NED, such as oscillons \cite{Cotaescu:2000lpp} 
and breathers \cite{Belinsky:1991aa}, which represent time-dependent localized structures.

In addition to these directions, 
it is worth considering whether nonsingular charged black hole solutions in NED may represent classical models of elementary particles. 
Dymnikova constructed spherically symmetric, nonsingular solutions in NED coupled to gravity, which resemble electrons in their finite energy and localized charge distribution \cite{Dymnikova:2021vkb}.
Although not concerned with nonsingular black holes,
Burinskii proposed a classical model of the Dirac electron based on the Kerr-Newman geometry, highlighting the agreement of its gyromagnetic ratio with that predicted by Dirac theory \cite{Burinskii:2005mm}.
These studies suggest that nonsingular spacetime geometries may offer useful analogies to particle-like behavior.
Moreover, in order to mimic the electromagnetic field sourced by a point charge,
nonsingular charged solutions must approach the Reissner--Nordstr\"om solution at the spatial infinity.
Some well-known nonsingular solutions, such as the Bardeen and Hayward black holes, deviate from the Coulombic behavior of Maxwell theory at large distance.
In contrast, the Fan--Wang black hole can exhibit Reissner--Nordstr\"om-like asymptotics under particular parameter choice, making it a more promising candidate for a particle-like configuration. 
It is also interesting to study the conditions to accomplish the suitable asymptotic behavior,
which will be our future work.

%%%%%%%%%%%%%%%%%%%%%%%%%%%%%%%%%%%%%%%%%%%%
%%%%%%%%%%%%%%%%%%%%%%%%%%%%%%%%%%%%%%%%%%%%
%%%%%%%%%%%%%%%%%%%%%%%%%%%%%%%%%%%%%%%%%%%%

%        << Acknowledgments >>

%%%%%%%%%%%%%%%%%%%%%%%%%%%%%%%%%%%%%%%%%%%%
%%%%%%%%%%%%%%%%%%%%%%%%%%%%%%%%%%%%%%%%%%%%
%%%%%%%%%%%%%%%%%%%%%%%%%%%%%%%%%%%%%%%%%%%%

\acknowledgments
R. S. was supported by JSPS KAKENHI Grant Number JP18K13541. 
S. T. was supported by JSPS KAKENHI Grant Number 21K03560.

%%%%%%%%%%%%%%%%%%%%%%%%%%%%%%%%%%%%%%%%%%%%
%%%%%%%%%%%%%%%%%%%%%%%%%%%%%%%%%%%%%%%%%%%%
%%%%%%%%%%%%%%%%%%%%%%%%%%%%%%%%%%%%%%%%%%%%

%                                << Reference>>

%%%%%%%%%%%%%%%%%%%%%%%%%%%%%%%%%%%%%%%%%%%%
%%%%%%%%%%%%%%%%%%%%%%%%%%%%%%%%%%%%%%%%%%%%
%%%%%%%%%%%%%%%%%%%%%%%%%%%%%%%%%%%%%%%%%%%%

\end{document}